\documentstyle[twocolumn,prl,psfig,aps]{revtex}
\begin{document}
\twocolumn[\hsize\textwidth\columnwidth\hsize\csname @twocolumnfalse\endcsname 
\draft
\begin{title}
{ Magnetic Quantum Dot: A Magnetic Transmission Barrier and Resonator }
\end{title}
\author{H.-S. Sim,$^1$\cite{first} G. Ihm,$^2$ N. Kim,$^3$ and K. J. Chang$^1$}
\address{$^1$ Department of Physics, Korea Advanced 
	Institute of Science and Technology, 
	Taejon 305-701, Korea}
\address{$^2$ Department of Physics, Chungnam National 
	  University, Taejon 305-764, Korea }
\address{$^3$ Quantum-functional Semiconductor Research 
Center, Dongguk University, Seoul 100-715, Korea}
\date{\today}
\maketitle
\begin{abstract}
\widetext
We study the ballistic edge-channel transport in quantum wires with 
a magnetic quantum dot, which is formed by two different magnetic 
fields $B^*$ and $B_0$ inside and outside the dot, respectively.
We find that the electron states located near the dot and
the scattering of edge channels by the dot
strongly depend on whether $B^*$ is parallel or antiparallel to $B_0$.
For parallel fields, two-terminal conductance as a function of
channel energy is quantized except for resonances, while,
for antiparallel fields, it is not quantized and 
all channels can be completely reflected in some energy ranges.
All these features are attributed to 
the characteristic magnetic confinements caused by nonuniform fields.
\end{abstract}


]

\narrowtext

Transport properties of two-dimensional electron gas (2DEG) in 
spatially nonuniform magnetic fields have attracted much attention.
Various magnetic structures such as magnetic dots \cite{McCord},
superlattices \cite{Carmona}, barriers \cite{Leadbeater}, and 
transverse steps \cite{Nogaret} were realized experimentally in 
nonplanar 2DEGs or by patterning ferromagnetic or superconducting 
materials.
Theoretically, it was shown that nonuniform magnetic fields can 
cause electron drifts \cite{Mints,Gu}, transmission barriers 
\cite{Matulis}, and commensurability effects \cite{Peeters}. 
Magnetic edge states, which exist along the boundary between two 
different magnetic domains, were proposed \cite{Sim,Reijniers1}
in the analogy with the conventional edge states \cite{Halperin} 
in quantum Hall systems, and their effects on magnetoresistance 
were reported experimentally \cite{Nogaret}.

The electron transport through quantum wires in strong magnetic 
fields can be well described by edge channels. 
When a local electrostatic modulation is applied additionally 
inside the wires, conductances can be still quantized and 
resonant reflections appear \cite{Jain,Takagaki}.
These interesting features can be modified when such a modulation 
is replaced by a magnetic one such as a magnetic quantum dot 
(or magnetic antidot) \cite{Sim,Reijniers1}, which is formed in 
2DEG by nonuniform perpendicular magnetic fields;
$\vec{B}=B^* \hat{z}$ within a circular disk with radius $r_0$,
while $\vec{B}=B_0 \hat{z}$ outside it.
The classical electron trajectories (see Fig. 1) scattered by a 
magnetic dot with $\gamma~[=B^*/B_0] < 0$ are very different from 
those for $\gamma > 0$ and those by an electrostatic dot or antidot,
indicating distinct edge-channel scatterings by local magnetic 
modulations from those by electrostatic ones.
The study of such a scattering mechanism is important to 
understanding the electron transport in magnetic structures and to 
suggesting future device applications.
However, to our knowledge, little attention has been paid to it
\cite{Matulis}.

In this Letter, we study the ballistic transport of conventional 
edge channels through quantum wires with a magnetic quantum dot.
The magnetic edge states near the dot and the two-terminal 
conductance $G(E_F)$ of the wires in the zero bias limit
are found to exhibit distinct features between two cases of 
$\gamma > 0$ and $\gamma < 0$, where $E_F$ is the Fermi energy.
For $\gamma > 0$, $G(E_F)$ is quantized 
and the dot behaves as a transmission barrier and a resonator, 
when the magnetic length inside the dot is smaller than $r_0$.
This feature results from the harmonic-potential-like magnetic 
confinements and is similar to that of electrostatic modulations.
On the other hand, for $\gamma < 0$, $G(E_F)$ is not quantized
when incident channels are scattered by the dot.
Moreover, for $\gamma < -1$, all incident channels can be 
completely reflected by the dot in some ranges of $E_F$, resulting 
in the plateaus of $G(E_F)=0$.
These interesting features for $\gamma < 0$ are due to the 
double-well and merged-well magnetic confinements 
caused by the field reversal at the dot boundary.
We also propose a calculational method for conductances,
based on the symmetric gauge and Green's function.

The dot is located in the middle of an infinitely long wire, 
whose potential is an infinite square well of width $L_y$ in the
transverse $y$-direction (see Fig. 1).
The distance between the dot and wire edge is $\Delta$.
The magnetic length and the Landau energy inside (outside) the dot 
are determined by $B^*$ ($B_0$) as
$l_{B^*}(j)=\sqrt{(2j+1)\hbar/(e|B^*|)}$ 
[$l_{B_0}(j)=\sqrt{(2j+1)\hbar/(eB_0)}$] and
$E^*(j) = (j+1/2)\hbar e |B^*| / m^*$
[$E_0(j) = (j+1/2)\hbar e B_0 / m^*$], respectively,
where $j$ = 0,1,2,... and $m^*$ is the effective mass.
We focus on the edge state transport regime, {\it i.e.}, 
$L_y \gg l_{B_0}(N-1)$, and ignore the effects of spin and disorder, 
where $N$ is the number of Landau levels below $E_F$ 
far away from the dot.

We first consider the case for $\Delta > l_{B_0}(N-1)$.
In this case, incident conventional edge channels do not interact 
with the dot, thus, $G(E_F)=NG_0$ (including the spin degeneracy) 
except for backward scatterings of channels by the resonant tunneling 
into the magnetic edge states of the dot, where $G_0 = 2e^2/h$.
To understand these magnetic edge states, we examine a magnetic dot 
in an infinite 2DEG.
When the dot center is located at $r=0$ in polar coordinates 
($r$,$\theta$), the eigenstates can be written as 
$\psi_{nm}(\vec{r}) = R_{nm}(r) e^{im\theta}$ for the symmetric gauge,
where $m$ is the angular momentum quantum number and
$n$ (=0,1,2,...) is the number of nodes in $R_{nm}(r)$.

The eigenstates can be classified by their radial locations.
A ($n$,$m<0$) state located far away from the dot interacts with $B_0$.
From the gauge invariance \cite{GAUGE}, its radial wave function is 
found to be the same as that of the ($n$,$m_{\rm eff}$) state 
in uniform fields $B_0$.
Here, $m_{\rm eff}=m-s$ and $s$ [$=(1-\gamma) \pi r_0^2 B_0 / \phi_0$] 
is the number of removed magnetic flux quanta (or additional ones for 
$s<0$) to form the magnetic dot in 2DEG where the uniform $B_0$ is 
already applied \cite{Sim}.
Then, $\psi_{nm<0}$ is located at $r_p(m_{\rm eff},B_0)$ with 
the eigenenergy $E_{nm}=E_0(n)$, enclosing $|m|$ flux quanta, where 
$r_p(m,B) = \sqrt{2|m|\hbar/(eB)}$.
On the other hand, $\psi_{nm}$'s near the dot interact with both 
$B_0$ and $B^*$, thus, $E_{nm}$'s deviate from $E_0(n)$.
They are magnetic edge states circulating along the dot boundary 
\cite{Sim} and cause resonant scatterings of conventional edge 
channels.
When $r_0 \gg l_{B^*}(n-1)$ and $|m|$ is small,
$\psi_{nm}$'s are located at $r_p(m,B^*)$ inside the dot with
$E_{nm} = E^*(n)$. 
Interestingly, for $\gamma<0$, $\psi_{nm}$'s with small 
$m < 0$ can be located also at $r_p(m_{\rm eff},B_0)$ outside
the dot.

The above features are clearly shown in Fig. 2.
In dimensionless units of $E_0(0) \rightarrow 1$ and 
$\sqrt{2} l_{B_0}(0) \rightarrow 1$, we calculate $E_{nm}$'s
from the radial part of the Schr$\rm\ddot{o}$dinger equation,
$[d^2/dr^2 + d/(rdr) + 2(E_{nm}-V_{\rm eff}(r))]R_{nm}(r)=0$.
Here, we define the {\it magnetic confinement} as the effective
potential $V_{\rm eff}$,
where $V_{\rm eff}(r) = (m/r + \gamma r)^2/2$ for $r<r_0$,
while $V_{\rm eff}(r) = (m_{\rm eff}/r + r)^2/2$ for $r>r_0$. 
For $\gamma > 0$, $V_{\rm eff}(r)$ is similar to the harmonic 
potential, which is the magnetic confinement in uniform fields.
Thus, $E_{nm}$'s vary monotonously from $E_0(j)$ at large $|m|$ 
[{\it i.e.}, $r_p(m_{\rm eff},B_0) \gg r_0$] 
to $E^*(j)$ at small $|m|$ [{\it i.e.}, $r_p(m,B^*) \ll r_0$],
where $j=n+(m+|m|)/2$.
Note that for small $r_0$ [$< l_{B^*}(j)$], $E_{nm}$ does not 
reach $E^*(j)$ at small $|m|$, as shown in Fig. 2(b).
For $\gamma > 1$, magnetic edge states circulate counterclockwise 
around the dot, while either clockwise or counterclockwise for 
$0 < \gamma < 1$.

For $\gamma < 0$, magnetic confinements are very different from the 
harmonic potential.
For $|m| < |\gamma| s_0$, $V_{\rm eff}$ is a double-well potential
with a barrier at $r_0$, where $s_0 = \pi r_0^2 B_0 / \phi_0$.
The barrier is high enough
to confine $\psi_{nm}$ only in one of the wells, if $r_0 \gg l_{B^*}$.
Then, for small $m<0$, the inner well allows energies $E^*(j_1)$ with 
$j_1 = |m|, |m|+1, |m|+2$,..., while the outer well allows
$E_0(j_2)$ with $j_2 = 0,1,2$,...
Thus, $\psi_{nm}$ with small $m<0$ can be located either inside or 
outside the dot, depending on $n$, as discussed before. 
This feature results in abrupt changes of $E_{nm}$'s from $E_0$ to 
$E^*$ [see Fig. 2(d)].
Note that in Fig. 2(c) the abrupt change appears only in the $n=0$ 
level, since $l_{B^*}(0) \approx r_0$.
For $|\gamma| s_0 \le m \le s+s_0$, the two wells in $V_{\rm eff}$
merge into a single well with a minimum at $r_0$ 
[see the dotted line in Fig. 2(e)].
The magnetic edge states in this merged well circulate 
counterclockwise along $r=r_0$ with snake-like classical motions.

Next, we study the scattering of conventional edge channels by the 
magnetic dot when $\Delta \le l_{\rm B_0}(N-1)$.
We calculate the transmission probability $T(\alpha)$ 
[$= G(\alpha)/G_0$] of incident channels as a function of 
dimensionless energy $\alpha$ [$= E_F/(2E_0(0))$] using the lattice 
Green's function \cite{Takagaki}, where a continuous 2DEG is 
approximated by a tight-binding square lattice with 
lattice constant $a$.
The vector potential is included as the Peierls' phase
factor [$exp(-ie / \hbar \int_l \vec{A} \cdot d \vec{l})$] 
in hopping matrix elements.  
While most previous studies have chosen the Landau gauge for this
approach, the symmetric gauge is essential in our work, which modifies
the approach \cite{METHOD}.

The behavior of $T(\alpha)$ can be classified by $\gamma$ (see Fig. 3).
For $\gamma > 0$, $T(\alpha)$ is quantized when $l_{B^*} < r_0$.
In this case, the magnetic confinements are similar to the harmonic 
potential. 
Thus, when edge channels pass the constriction between the dot and 
wire edge, they are still well confined near the wire edge, without 
the interaction with those in the opposite edge, resulting in the 
quantization of $T(\alpha)$.

For $\gamma > 1$, $T(\alpha)$ is smaller than that of the 
uniform-field case with $\gamma=1$, as in the case of electrostatic 
antidots \cite{Takagaki}.
This feature results from the fact that some incident edge channels 
are reflected by the dot due to the magnetic energy $E^*$ larger 
than $E_0$.
As $\Delta$ decreases, the transition energy $E_t(j)$, where $T$ 
changes from $j$ to $j+1$, increases from $E_0(j)$ to $E^*(j)$ 
and the number of resonances decreases [see Figs. 3(a) and (c)],
because the magnetic edge states are confined in a narrower region.
For $0 < \gamma < 1$, $T(\alpha)$ is the same as that for $\gamma=1$ 
except for resonant dips.
In this case, since $E^* < E_0$, the dot does not reflect any 
incident edge channels and binds electrons, as in the case of 
electrostatic dots; thus, the magnetic dot behaves as a resonator.
More resonances occur as $\gamma$ decreases from 1 and $r_0$ 
increases.

The features for $\gamma < 0$ are very different from those for 
$\gamma > 0$ and those by electrostatic modulations.
For $-1 < \gamma < 0$, $T(\alpha)$ is not quantized and smaller 
than that of the uniform-field case, although $E^* < E_0$, 
in contrast to the case of $0 < \gamma < 1$.
For $\gamma < -1$, $T(\alpha)$ is not quantized.
Moreover, when $\Delta \approx l_{B_0}(0)$, all incident channels 
are completely reflected in some ranges of $\alpha$, except for 
resonances, so that $G(\alpha)$ oscillates between 0 and $G_0$ with 
the plateaus of $G=0$, in marked contrast to the magnetic dot with 
$\gamma > 1$.

The features for $\gamma < 0$ result from the double-well and 
merged-well magnetic confinements, which are caused by the field 
reversal.
To understand this behavior, we imitate the region near the dot by 
a magnetic step \cite{Gu} in an infinite square well $U(y)$ with 
width $L_y$, which is divided into three strips by different 
magnetic fields;
$B=B^*$ in the middle strip ($|y| < r_0$),
while $B=B_0$ in the upper ($y > r_0$) and lower ones ($y < -r_0$).
Its eigenstates can be written as $e^{ikx}Y_k(y)$, and 
$V_{\rm eff}(y,k)$ is defined in a similar way to the dot case;
$V_{\rm eff}(y,k) = \hbar^2 \{k+F(y)/l_{B_0}^2\}^2/(2m^*) + U(y)$,
where $F(y)$ is $y+r_0(\gamma-1)y/|y|$ for $|y| > r_0$ and
$\gamma y$ for $|y| < r_0$.
In Fig. 4, $V_{\rm eff}(y)$'s and the calculated energy levels 
$E(k>0)$'s are shown; 
note that $E(k<0)=E(|k|)$. 
The states near $y = \pm L_y/2$ correspond to the current-carrying
conventional edge states, while those near $y= \pm r_0$
to the magnetic edge states circulating around the dot.
The triple wells [solid and dashed lines in Figs. 4(a)-(b)] 
correspond to the double-well magnetic confinements of the magnetic 
dot, while the double wells (dotted lines) to the merged-well ones.

For $-1 < \gamma < 0$, as $\Delta$ ($= L_y/2 - r_0$) decreases,
the edge states with a given energy 
near $y = - L_y/2$ are determined by $V_{\rm eff}$
with smaller $k > 0$, which has a smaller barrier at $y=-r_0$,
due to the wire confinement [see Fig. 4(a)].
When $\Delta \approx l_{B_0}$, the barrier is so small that the 
states near $y = -L_y/2$ can be easily extended to the middle or 
upper strip.
The same behavior arises in the case of the magnetic dot.
When $\Delta \approx l_{B_0}$, the conventional edge channels can 
interact with the double-well magnetic confinement with a small 
barrier, so that they are extended in the transverse direction. 
Then, the left-going channels easily interact with the right-going 
ones, thus, the conductance is not quantized, well corresponding to
the classical trajectories in Fig. 1.

For $\gamma < -1$ and $k>0$, 
when $l_{B_0} \ll \Delta < |\gamma| r_0$, 
the states confined in the local minimum at $y=L_y/2$ of the triple 
wells are the conventional edge states [see Fig. 4(b)].
Their energies are much higher than $E_0$ at $k=0$ 
and satisfy $dE/dk < 0$.
As $\Delta$ decreases, the well at $y=L_y/2$ becomes narrower, so 
that these states have higher energies and begin to be mixed with 
the magnetic edge states near $r=r_0$, resulting in level splitting.
The number of the pure conventional edge channels near $y=L_y/2$ is 
$\sim M$, where $M$ is the largest number satisfying 
$2l_{B_0}(M-1) < \Delta$.
In Fig. 4(c), the energy levels of two pure channels are shown; 
note that the levels of channels near $y=-L_y/2$ do not appear 
because of their very high energies.
When $\Delta \approx l_{B_0}(0)$, there exist no pure conventional
edge channels.
In this case, eigenstates are classified into those with $dE/dk = 0$ 
inside the middle strip, those with $dE/dk > 0$ caused by the merged 
wells at $y=r_0$, and those with $dE/dk < 0$ which are the mixed 
ones of the magnetic and conventional edge states due to the 
triple wells with a small barrier at $r_0$ [see Fig. 4(d)].
Only the third ones propagate in the same direction as the 
conventional edge states, and they are not allowed in some energy 
ranges above $E^*$'s because of the state mixing.
This feature indicates that all conventional edge channels can not 
pass the constriction between the magnetic dot and wire edge, 
{\it i.e.}, $G(E_F)=0$, in some ranges above $E^*$'s. 
The plateaus of $G(E_F)=0$ appear in wider energy ranges
for larger $|\gamma|$, smaller $E_F$, and smaller $\Delta$.
The resonant peaks in the ranges of $T(\alpha)=0$ in Fig. 3(b) 
result from the snake magnetic edge states in the merged-well 
magnetic confinements.

Magnetic modulations with an order of $\sim 1$ T 
($\alpha \sim 2$ for electron density of $10^{11}$ $cm^{-2}$ and 
$l_{B_0}(1) \sim 45$ $nm$)
have been realized for both $\gamma > 0$ and $\gamma < 0$ 
in nonplanar 2DEGs \cite{Leadbeater},
while those of $\sim 0.1$ T 
($\alpha \sim 20$ and $l_{B_0}(19) \sim 500$ $nm$)
in magnetic steps \cite{Nogaret}.
When a magnetic dot or a finite step is formed by such modulations,
our findings can be observed, because the magnetic confinements for 
$\gamma < 0$ ($\gamma > 0$) still form the double wells or merged 
wells (harmonic-like potentials).
The modulations in Ref.~\cite{Leadbeater} can be considered
as a magnetic dot with very large $r_0$ and $\Delta=0$.
We propose that the constrictions between a magnetic dot and wire 
edges behave as a {\it magnetic quantum point contact}.
The conductance in this new device with $\gamma > 1$ is similar to 
that in electrostatic contacts \cite{Wees}, 
while for $\gamma < -1$, it can be very different, 
showing a switching behavior with the plateaus of $G(E_F)=0$.

In conclusion, we find that
a magnetic quantum dot in a quantum wire behaves as
a characteristic transmission barrier and a resonator.
The double-well and merged-well magnetic confinements caused by the 
field reversal at the dot boundary result in distinct magnetic edge 
states and transport properties, such as the nonquantized conductance 
and the plateaus of $G=0$.
The magnetic confinements are important to understanding
the electronic and transport properties of other magnetic structures.

We are grateful to Prof. F. M. Peeters for sending copies of his works.
This work was supported by the QSRC at Dongguk University.

\begin{figure}
\centerline{\psfig{figure=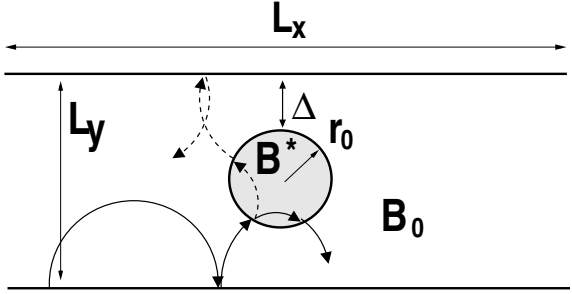,
height=,width=0.45\textwidth}}
\caption{Schematic diagram of a quantum wire with a magnetic quantum dot.
Solid (dashed) arrows represent the classical electron trajectories
for $B^*/B_0 > 0$ ($B^*/B_0 < 0$).}
\label{fig1}
\end{figure}
\begin{figure}
\centerline{\psfig{figure=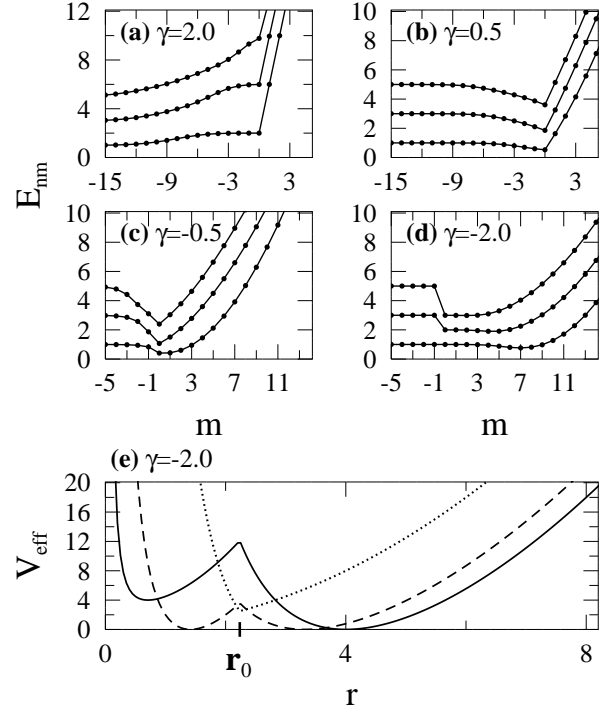,
height=,width=0.45\textwidth}}
\caption{(a)-(d) $E_{nm}$'s and (e) $V_{\rm eff}(r,m)$'s 
for $s_0$ (= $\pi r_0^2 B_0 / \phi_0$) $=5$ and some $\gamma$'s. 
In (e), $m$= -1 (solid), 4 (dashed), 15 (dotted) are chosen.
The energy unit is $E_0(0)$.}
\label{fig2}
\end{figure}
\begin{figure}
\centerline{\psfig{figure=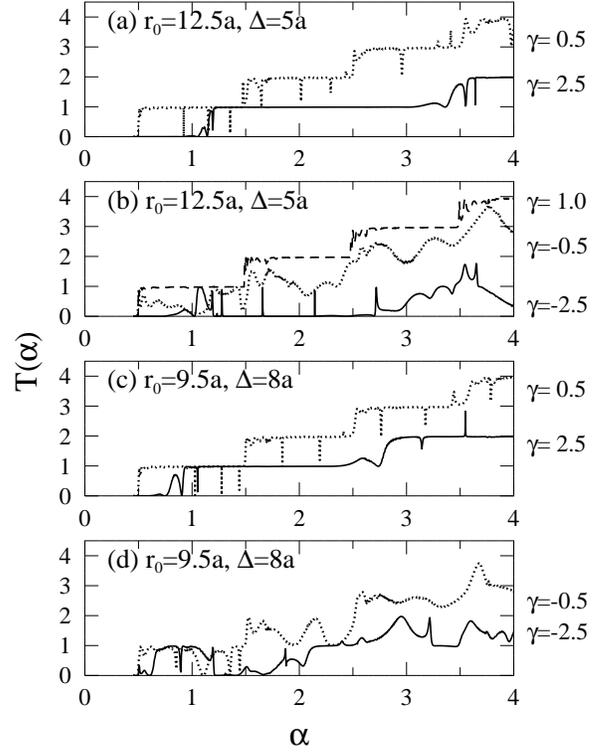,
height=,width=0.45\textwidth}}
\caption{$T(\alpha)$ for some $\gamma$'s and $r_0$'s. 
For all cases, $L_y = 35a$ and $l_{B_0}(0) = 5a$.}
\label{fig3}
\end{figure}
\begin{figure}
\centerline{\psfig{figure=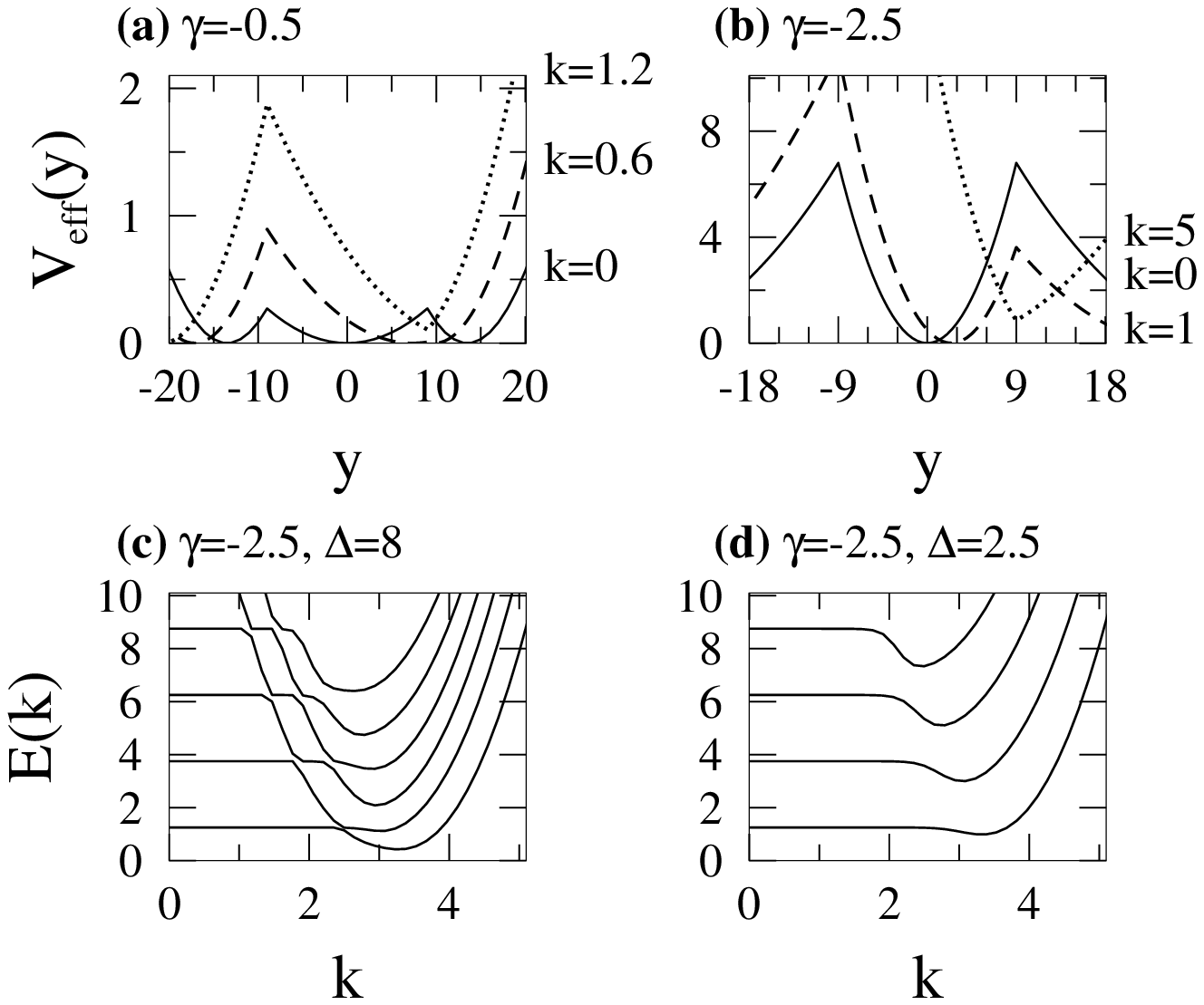,
height=,width=0.45\textwidth}}
\caption{(a)-(b) $V_{\rm eff}(y,k)$'s and (c)-(d) $E(k)$'s for 
magnetic steps.
The energy unit is $2E_0(0)$ while the length unit is arbitrary.
For all cases, $l_{B_0}(0) = 2.47$ and $r_0 = 9$.}
\label{fig4}
\end{figure}

\begin{references}
\bibitem[\dagger]{first} Present address: Max-Planck-Institute for
the Physics of Complex Systems, D-01187 Dresden, Germany.
\bibitem{McCord} M. A. McCord and D. D. Awschalom,
Appl. Phys. Lett. {\bf 57}, 2153 (1990).
\bibitem{Carmona} H. A. Carmona {\it et al.},
Phys. Rev. Lett. {\bf 74}, 3009 (1995);
P. D. Ye {\it et al.},
Phys. Rev. Lett. {\bf 74}, 3013 (1995);
A. Nogaret {\it et al.},
Phys. Rev. B {\bf 55}, R16037 (1999).
\bibitem{Leadbeater} M. L. Leadbeater {\it et al.},
J. Phys.: Condens. Matter {\bf 7}, L307 (1995);
Phys. Rev. B {\bf 52}, R8629 (1995).
\bibitem{Nogaret} A. Nogaret, S. J. Bending, and M. Henini,
Phys. Rev. Lett. {\bf 84}, 2231 (2000).
\bibitem{Mints} R. G. Mints, JETP Lett. {\bf 9}, 387 (1969);
J. E. M$\rm\ddot{u}$ller,
Phys. Rev. Lett. {\bf 68}, 385 (1992).
\bibitem{Gu} B.-Y. Gu {\it et al.},
Phys. Rev. B {\bf 56}, 13434 (1997);
J. Reijniers and F. M. Peeters, Phys. Rev. B {\bf 63}, 165317 (2001).
\bibitem{Matulis} A. Matulis, F. M. Peeters, and P. Vasilopoulos,
Phys. Rev. Lett. {\bf 72}, 1518 (1994);
J. Reijniers, F. M. Peeters, and A. Matulis,
Physica E {\bf 6} 759 (2000).
\bibitem{Peeters} F. M. Peeters and P. Vasilopoulos,
Phys. Rev. B {\bf 47}, 1466 (1994).
\bibitem{Sim} H.-S. Sim {\it et al.},
Phys. Rev. Lett. {\bf 80}, 1501 (1998).
\bibitem{Reijniers1} J. Reijniers, F. M. Peeters, and A. Matulis,
Phys. Rev. B {\bf 59}, 2817 (1999).
\bibitem{Halperin} B. I. Halperin,
Phys. Rev. B {\bf 25}, 2185 (1982);
M. B$\rm\ddot{u}$ttiker,
Phys. Rev. B {\bf 38}, 9375 (1988).
\bibitem{Jain} J. K. Jain and S. A. Kivelson,
Phys. Rev. Lett. {\bf 60}, 1542 (1988).
\bibitem{Takagaki} Y. Takagaki and D. K. Ferry,
Phys. Rev. B {\bf 48}, 8152 (1993).
\bibitem{GAUGE} Let us consider a state with $m^\prime < 0$ 
[$\psi_{m^\prime} = e^{im^\prime\theta}R_{m^\prime}(r)$]
in the uniform field $B_0$.
This state is located at $r_p(m^\prime,B_0)$ and encloses $|m^\prime|$ 
flux quanta.
If $s$ flux quanta are removed inside $r_0$ ($\ll r_p$),
$\psi_{m^\prime}$ can be written as 
$e^{i(m^\prime+s)\theta}R_{m^\prime}(r)$ due to the gauge invariance
and then rewritten as $e^{im\theta}R_{m_{\rm eff}}(r)$.
\bibitem{METHOD}
To make $\vec{A}=0$ in leads \cite{Takagaki}, which are the wire regions 
of $|x|>L_x/2$, we choose that $B=0$ for $r>r_1$ ($\gg r_0$) and
$\vec{A}=0$ for $r>r_2$ ($\sim L_x/2$), where $r_1=38a$ and $L_x=4000a$.
This choice of $\vec{A}$ does not affect our results only when 
$L_y / L_x \ll 1$; otherwise, the Peierls' phase is counted incorrectly.
To imitate the continuum limit,
the condition $|2\pi B a^2 / \phi_0| \ll 1$ should be satisfied.
\bibitem{Wees} B. J. van Wees {\it et al.},
Phys. Rev. Lett. {\bf 60}, 848 (1988);
D. A. Wharam {\it et al.},
J. Phys. C {\bf 21}, L209 (1988).
\end{references}
\end{document}